\documentstyle[aps]{revtex}
\begin{document}
\draft
\def\lsim{\lower.5ex\hbox{$\; \buildrel < \over \sim \;$}}
\def\gsim{\lower.5ex\hbox{$\; \buildrel > \over \sim \;$}}
\title{Semi-analytical Solution of Dirac equation in Schwarzschild Geometry}
\author  {Banibrata Mukhopadhyay\\
     and\\
    Sandip K. Chakrabarti\\}

\vskip0.5cm
\address{Theoretical Astrophysics Group\\
S. N. Bose National Centre For Basic Sciences,\\
JD Block, Salt Lake, Sector-III, Calcutta-700091\\} 	
\thanks{e-mail:  bm@boson.bose.res.in  \& chakraba@boson.bose.res.in\\}
\maketitle
\vskip0.3cm
\setcounter{page}{1}
\noindent{To appear in Classical and Quantum Gravity}
\def\ch{\lower-0.55ex\hbox{--}\kern-0.55em{\lower0.15ex\hbox{$h$}}}
\def\lh{\lower-0.55ex\hbox{--}\kern-0.55em{\lower0.15ex\hbox{$\lambda$}}}	

\begin{abstract}

Separation of the Dirac equation in the spacetime around a Kerr black hole
into radial and angular coordinates was done by Chandrasekhar in 1976. 
In the present paper, we solve the radial equations
in a Schwarzschild geometry semi-analytically 
using Wentzel-Kramers-Brillouin approximation (in short WKB) 
method. Among other things, we
present analytical expression of 
the instantaneous  reflection and transmission coefficients
and the radial wave functions of the Dirac particles. 
Complete physical parameter space was divided
into two parts depending on the height of the potential well 
and energy of the incoming waves.
We show the general solution for these two regions.
We also solve the equations by a Quantum Mechanical approach, in which
the potential is approximated by a series of steps and found
that these two solutions agree.
We compare solutions of different initial parameters
and show how the properties of the scattered wave depend
on these parameters.

\end{abstract}


\section*{I. INTRODUCTION}

The spacetime around an isolated black hole
is flat and Minkowskian at a large distance
where usual quantum mechanics is applicable, while the spacetime
closer to the singularity is curved and no satisfactory
quantum field theory could be developed as yet. However,
occasionally, it is useful to look into an intermediate
situation when a weak perturbation (due to, say, gravitational,
electromagnetic or Dirac waves) originating from
infinity scatters from a black hole.
The resulting wave is partially transmitted into the
black hole through the horizon and partially scatters off from
it to infinity. In the linearized (`test field') approximation
this problem has been attacked in the past by several authors [1-4].
These methods are mostly numerical and most of the solutions 
obtained so far is for particles of integral spin  only.
Chandrasekhar [3-4] separated the Dirac equation in Kerr geometry into
radial and angular parts. These works were extended to
other spacetimes, such as in Kerr-Newman geometry [5], and  around
dyon black holes [6]. Subsequently, Chakrabarti [7] solved the 
angular part of the Dirac equation in Kerr geometry and gave
the eigenvalues of the equation.  These and the present
works mostly concern scattering off tiny black holes and thus
changing the incoming solution appreciably into an outgoing solution.
Scattering  effects from larger black holes could be studied by phase
shift analysis and these has also been done recently [8].

In the present paper, we attack a simpler problem to have a `feel' for the
complete solution when the black hole is non-rotating.
In the next Section, we present the basic equations. In \S 3, we
classify the parameter space in terms of the physical and unphysical
regions and present the method we adopt to solve the equations.
In \S 4, we present a complete solution. In \S 5, we present
solutions using a classical method in which the potential is
approximated by a series of steps and then compare solutions
of these two methods. In \S 6, we also compare solutions
of various parameters and show how a Schwarzschild black
hole distinguishes incoming particles of various masses.
Finally, in \S 7, we draw our conclusions.

\section*{II. BASIC EQUATIONS OF THE PROBLEM}

Following Chandrasekhar [4], the radial part of the Dirac equation
is easily reduced into a Schr\"odinger like equation. However, because
the spin-spin coupling term is absent in the Schwarzschild geometry, 
the radial equation is much simpler to deal with. The eigenvalue of 
the angular equation for spin $\pm 1/2$
is trivially obtained as $(l+1/2)^2$ [7, 9-10] where $l$ is the orbital
quantum number. In what follows, we choose $l = 1/2$ throughout for concreteness.
This eigenvalue turns out to be the separation constant  $\lambda$ of the original Dirac 
equation [4]. Here we solve the equation for one possible value
of separation constant $\lambda$ (for $l=1/2$, $\lambda$ is unity).
In future we plan to explore the nature of the solutions for other orbital 
quantum numbers.

Presently, we need to solve only the following coupled radial equations [4]:

$$
\Delta^{1 \over 2} { \cal D}_{0} {R}_{-{1 \over 2}} = 
( 1 + i m_p  r) \Delta^{1 \over 2} R_{+{1 \over 2}} ,
\eqno{(1)}
$$

$$ 
\Delta^{1 \over 2} {\cal D}_{0}^{\dag}\Delta^{1 \over 2} 
{R}_{+{1 \over 2}} = ( 1 - i m_p  r)  {R}_{-{1 \over 2}} ,
\eqno{(2)}
$$

where,
$$
{\cal D}_n = \partial_{r} + \frac {i K} {\Delta} + 2n \frac{(r-M)} {\Delta},
$$ 

$$
\Delta = r^2 - 2Mr,
$$

$$
K=r^2\sigma.
$$
Here $n$ = integer, $\sigma$ = frequency of incoming Dirac wave,
$M$ = mass of the black hole, $m$ = azimuthal quantum number, $m_p$ = rest 
mass of the Dirac particle; p indicates particle,  
$R_{+{1 \over 2}} (R_{-{1 \over 2}})$ = radial 
wave function for  spin up (down) particles.
${\cal D}_{0}^{\dag}$ is the complex conjugate operator.
It is to be noted that the dimensionless unit is chosen, so that $G=\ch=c=1$.
The radial equation here is in coupled form. We can decouple it and express
the equation either in terms of spin up or spin down wave function.
However, it is more convenient to follow
Chandrasekhar's [4] approach by which the basis was changed along with the independent
variable $r$. That way, the coupled equation was reduced into two independent
one dimensional wave equations since they are easier to solve. 

We first define
$$
{r}_* = r + 2M log \left| r - 2M \right|,
\eqno{(3)}
$$
where, $r > r_{+} (= 2M)$, 
$$
\frac{d}{d{{r}}_*} = \frac{\Delta}{r^2} \frac{d}{dr},
\eqno{(4)}
$$
and choose $\Delta^{1 \over 2} R_{+ {1 \over 2}} = P_{+ {1 \over 2}}$,
$R_{- {1 \over 2}} = P_{- {1 \over 2}}$.

In terms of $r_{*}$, the operators take the form: 
$$
{\cal D}_{0} = {r^{2} \over \Delta} ({d \over dr_*} + i \sigma)
$$
and 
$$
{\cal D}^{\dagger}_{0}={r^2 \over \Delta} ({d \over dr_{*}} - i\sigma) .
$$

We choose $\theta = {\rm tan}^{-1} (m_p r)$ which yields,
$$
{\rm cos} \theta = \frac{1}{\surd(1 + m_p^2 r^2)}, \ \ \ {\rm sin} \theta = \frac{m_p r} {\surd(1 + m_p^2r^2)}
$$
and 
$$
(1 \pm i m_p r) = exp ({\pm i \theta}) \surd (1 + m_p^2 r^2).
$$

Following exactly Chandrasekhar's [4] approach  we write 
$$
P_{+ {1 \over 2}} = \psi_{+ {1 \over 2}}  exp\left[-{1 \over 2}
i {\rm tan}^{-1} (m_p r)\right]
\eqno{(5)}
$$ 
and
$$
P_{- {1 \over 2}} = \psi_{- {1 \over 2}}  exp\left[+{1 \over 2} 
i  {\rm tan}^{-1} (m_p r)\right] .
\eqno{(6)}
$$

Finally, a choice of  $\hat{r}_* = r_{*} + {1 \over 2\sigma} {\rm tan}^{-1}
(m_p r)$ yields $d{\hat r}_* = \left(1 + {\Delta \over r^2} 
{{ m_p} \over 2 \sigma} {1 \over {1 + m_p^2 r^2}}\right)dr_*$.

With these definitions, the differential equations  (1-2) are re-written as
$$
\left(\frac{d} {d{\hat r}_*} - W\right) Z_+ = i \sigma Z_-
\eqno{(7a)}
$$
and
$$
\left(\frac{d} {d{\hat r}_*} + W\right) Z_- = i \sigma Z_+ ,
\eqno{(7b)}
$$
where, $Z_{\pm} = \psi_{+ {1 \over 2}} \pm \psi_{-{1 \over 2}}$, and
$$
W = \frac{\Delta^{1 \over 2} (1 + m_p^2 r^2)^{3/2} } 
{ r^2 (1 + m_p^2 r^2) + m_p \Delta/2\sigma}.
\eqno{(8)}
$$
One important point to note: the transformation of spatial coordinate $r$ to
$r_{*}$ (and ${\hat{r}}_{*}$) is taken not only for mathematical simplicity
but also for a physical significance.
When $r$ is chosen as the radial coordinate, the decoupled equations
for independent waves show diverging behaviour.
However, by transforming those in terms of $r_{*}$ (and ${\hat{r}}_{*}$) 
we obtain well
behaved functions. The horizon is shifted from $r=r_+$ to ${\hat {r}}_{*}= -\infty$.

>From the above set of equations, we readily obtain a pair of independent 
one-dimensional wave-equations, 
$$
\left(\frac{d^2} {d {\hat r}_*}^2 + \sigma^2\right) Z_\pm = V_\pm Z_\pm ,
\eqno{(9)}
$$
where, 
$$
V_{\pm} = W^{2} \pm {dW \over d\hat{r}_{*}}
\eqno{(10)}
$$
$$
\ \ \ ={{\Delta^{1 \over 2}(1 + m_p^{2} r^{2})^{3/2}} \over {[ r^{2}(1 + m_p^{2} 
r^{2}) +  m_p \Delta/2 \sigma]^{2}}}[\Delta^{1 \over 2}(1 + m_p^{2} r^{2})^{3/2}
\pm \left((r-M)(1 + m_p^{2} r^{2}) + 3m_p^{2} r \Delta\right)]
$$
$$
\mp {{\Delta^{3 \over 2}(1 + m_p^{2} r^{2})^{5/2}} \over {[ r^{2}(1 + m_p^{2}
r^{2}) +  m_p \Delta/2 \sigma]^{3}}}[2r(1 + m_p^{2} r^{2}) + 2 m_p^{2} r^{3}  +
 m_p (r-M)/\sigma].
\eqno{(11)}
$$

\section*{III. PARAMETER SPACE AND METHOD TO SOLVE EQUATIONS}

We obtain solutions by employing WKB [11-12] method 
and then imposing strict boundary conditions 
on the horizon, so that the reflection coefficient is zero 
and transmission coefficient is unity at the horizon. 
After establishing the general solution, we present here the 
solution of eq. (9) for two sets of parameters as illustrative examples. 

It is advisable to choose the parameters in such 
a way that there is a significant interaction between the particle and the 
black hole. This is possible when the Compton wavelength of the incoming 
wave is of the same order as the Schwarzschild radius of the black hole, i.e.,
$$
\frac{2GM}{c^2} \sim \frac{\ch}{{m_p}c},
$$
here, we are choosing $G = \ch = c = 1 $, so 
$$
m_p \sim \frac{1}{2M}.
$$
Again, in the case of Schwarzschild geometry, frequency of the incoming
particle (or, wave) will be of the same order as inverse of time. So,
$$
\frac{c^3}{2GM} \sim \sigma.
$$
Using the units as before, one can write,
$$
m_p \sim \sigma \sim {(2M)}^{-1}.
\eqno{(12)}
$$
In principle, however, one can choose any values of $\sigma$ and $m_p$ 
for a particular black hole and the corresponding solution is possible but we
shall concentrate the region of parameter space where the solution is
expected to be interesting as pointed out above, namely, region close to $m_p=\sigma$.
In Fig. 1a, we draw this line. The parameter space is spanned by the frequency
$\sigma$ and the rest mass of the incoming particle $m_p$.
It is clear that $50\%$ of total parameter space where $\sigma<m_p$
is unphysical, and one need not study this region. 
Rest of the parameter space ($\sigma>m_p$) is divided into two 
regions -- I: $E>V_{m}$ and II: $E<V_{m}$, where $V_{m}$ is the maximum of the
potential. While in Region I, the wave is  
{\it locally} sinusoidal because the wave number $k$ is real 
for the entire range of ${\hat r}_*$. In Region II, on the other hand,
the wave is decaying in some region when $E <V$, i.e., where
the wave `hits' the potential barrier and in the rest of the
region, the wave is propagating. We shall show
solutions in these two regions separately. In
Region-I whatever be the physical parameters, energy of the particle
is always greater than the potential energy and WKB approximation is
generally valid in the whole range (i.e. $\frac{1}{k^2}\frac{dk}
{d\hat{r}_*} << 1$). In cases of Region-II, energy of the particle 
is always less than the maximum height of potential barrier. Thus, at 
two points (where, $k=0$) total energy matches with the potential energy and 
in the neighbour of those two points WKB approximate method is not valid. 
They have to be dealt separately. In Fig. 1b, we show
contours of constant  $w_{max}=$max($\frac{1}{k^2}\frac{dk}{d\hat{r}_*}$) 
for a given set  ($\sigma, m_p$) of parameters. The labels 
show the actual values of $w_{max}$. Clearly, except for parameters
{\it very close} to the boundary of Regions I and II,
WKB approximation is safely valid for any value of $\hat{r}_*$. One has to
employ different method (such as using Airy Functions, see below)
to find solutions in this region.

\section*{IV. The complete solution}
\subsection*{Solutions of Region I}

In this region, for any set of parameter, energy of the particle is
always greater than the corresponding potential energy. We first 
re-write equation (9) as,
$$
{d^{2}Z_{+} \over d{\hat{r}}_{*}^{2}} + \left(\sigma^{2} - V_{+}\right) 
Z_{+} = 0.
\eqno{(13)}
$$
This is nothing but Schr\"odinger equation corresponding to the
total energy of the wave $\sigma^2$. This can be solved by 
regular WKB method [11-12]. Let, 
$$
k ({\hat r}_*) = \surd\left(\sigma^{2} - V_{+}\right),
\eqno{(14)}
$$
$$
u ({\hat r}_*) = \int k({\hat r}_*) d {\hat r}_* + {\rm constant}.
\eqno{(15)}
$$

Here, $k$ is the wavenumber of the incoming wave and $u$ is the {\it Eiconal}.
The solution of the equation (13) is,
$$
Z_{+} = \frac{A_{+}} {\surd k} exp (i u) + \frac {A_{-}} {\surd k} exp (- i u) .
\eqno{(16)}
$$
with
$$
A_+^2+A_-^2 = k.
\eqno{(17)}
$$
In this case all along $\sigma^{2} > V_{+}$ and also ${1 \over k}
{dk \over d{\hat{r}}_{*}} << k$, so WKB approximation is generally valid in the whole region.
The quantity $\frac{1}{k^2} {dk \over d{\hat{r}}_{*}}$ falls off rapidly with distance.
Thus, WKB is strictly valid at long distance only.

It is clear that a standard WKB solution where  $A_+$ and $A_-$ are kept constants throughout
should not be accurate, since the physical inner boundary condition on the horizon
must be that the reflected component is negligible there.
Thus WKB approximation requires a slight modification in which the
spatial dependence of $A_\pm$ is allowed. On the other hand,
at a large distance, where WKB is strictly valid, $A_+$ and $A_-$
should tend to be constants, and hence their difference is also a constant:
$$
A_+-A_-=c.
\eqno{(18)}
$$
Here, $c$ is determined from the WKB solution at a 
large distance. This along with (17) gives,
$$
A_\pm (r) = \pm {c \over 2} + {\sqrt{[2k(r) - c^{2}]} \over 2} .
\eqno{(19)}
$$
This spatial variation, strictly valid at large distances only,
should not be extendible to the horizon without
correcting for the inner boundary condition. These values
are to be shifted by, say, $A_{\pm h}$ respectively,
so that on the horizon one obtains physical $R$ and $T$.
We first correct reflection coefficient on the horizon as follows:
Let $A_{-h}$ be the value of $A_-$ on the horizon (see, equation (19)),
$$
A_{-h}= - {c \over 2} + {\sqrt{[2k(r_+) - c^{2}]} \over 2}.
$$
It is appropriate to use ${\cal A}_-=A_- - A_{-h}$  rather than $A_-$ since
${\cal A}_-$ vanishes at $r=r_+$. 

Incorporating these conditions, the solution (16) becomes,
$$
Z_{+} = \frac{{\cal A}_+} {\surd q} exp (i u) + 
\frac {{\cal A}_+} {\surd q} exp (- i u) .
\eqno{(20)}
$$
with the usual normalization  condition
$$
{\cal A}_+^2+{\cal A}_-^2 = q .
\eqno{(21)}
$$
where, ${\cal A}_+=A_+-A_{+h}$.
Here, $q$ is to be determined self-consistently by equating the asymptotic behaviour
of this reflection coefficient with that obtained using WKB method.
This $q$ in turn is used to compute ${\cal A}_+=A_+-A_{+h}$, and therefore
the transmission coefficient $T$ from eq. (21).
In this way, normalization of $R+T=1$ is assured.

Normalization factor $q \rightarrow k$ as $\hat{r}_{*} \rightarrow \infty$
and the condition $\frac{1}{q}\frac{dq}{d\hat{r}_{*}} << q$ is found
to be satisfied whenever $\frac{1}{k}\frac{dk}{d\hat{r}_{*}} << k$ is satisfied.
This is the essence of our modification of the WKB. In a true WKB, $A_\pm$ are
constants and the normalization is with respect to a (almost) constant $k$.
However, we are using it as if WKB is instantaneously valid everywhere.
Our method may therefore be called `Instantaneous' WKB approximation 
or IWKB for short. Using the new notations, the instantaneous values (i.e., local values)
of the reflection and transmission coefficients are given by (see, eq. 20),
$$
R= \frac {{\cal A}_-^2} { q}
\eqno{(22a)}
$$
$$
T= \frac {{\cal A}_+^2} { q} .
\eqno{(22b)}
$$
Determination of $A_{+h}$ is done by enforcing $R$ obtained from eq. (22a)
as the same as that obtained by actual WKB method at infinity.

To be concrete, we choose one set of parameters from Region I. (A large
number of solutions are compared in \S 6 below.) Here, total energy of
the incoming particle is greater than the potential barrier height for all
values of $\hat{r}_*$. We use mass of the
black hole, $ M = 1$; mass of the particle, $ m_p = 0.8$, orbital quantum
number, $l =  {1 \over 2}$, azimuthal quantum number, $m = - {1 \over 2}$,
frequency of the incoming wave, $ \sigma = 0.8$.

>From eq. (9) we observe that there are two wave equations for
two potentials $V_{+}$ and $V_{-}$. The nature of potentials
are shown in Fig. 2. It is clear that potentials
$V_\pm$ are well behaved. They are monotonically decreasing as the
particle approaches the black hole, and the total energy chosen
in this case ($\sigma^2$) is always higher compared to $V_\pm$.
For concreteness, we solve using potential $V_{+}$.
Similar procedure can be adopted using potential $V_{-}$ to compute $Z_-$ and
its form would be
$$
Z_{-} = \frac{A'_{+}-A'_{+h}} {\surd q'} exp (i u') - \frac {A'_{-}-A'_{-h}} 
{\surd q'} exp (- i u') .
\eqno{(20')}
$$
Note the occurrence of the negative sign in front of the reflected wave. This is to
satisfy the asymptotic property of the wave functions which must conserve 
the Wronskian [4]. 
Since the coefficients should not change sign between infinity and the
horizon (as that would tantamount to having zero amplitude, i.e.,
unphysical, absence of either the forward or the backward component) 
the same sign convention is followed throughout the space. Local values of 
the reflection and transmission coefficients could also be calculated 
in the same manner. In the solution (eq. $20$ and $20'$),
first term represents the incident wave and the second term 
represents the reflected wave. 

In Fig. 3  we show the nature of $V_{+}$ (solid curve), $k$ (dashed curve)
and $E (= \sigma^2)$ (short-dashed curve). The difference of $E$ and  $V_+$
and therefore $k$ goes up as the particle approaches the black hole.

In Fig. 4, variation of `local' reflection and transmission coefficients are shown. It 
is observed that as matter comes close to the black hole,
the barrier height goes down. As a result, the penetration probability
increases resulting in the rise of the transmission coefficients. At the
same time, the reflection coefficient tends to be zero. 
It is to be noted, that, strictly speaking, the terms 
`reflection' and `transmission' coefficients are traditionally
defined with respect to the asymptotic values. 
The spatial dependence that we show are to be interpreted  as the 
instantaneous values. This is consistent with the spirit of IWKB 
approximation that we are using.

The behaviour of the solutions with $V_-$ is not very different from what were shown 
in Figs. (3-4) except in a region very close to the black hole horizon 
where $V_+$ and $V_-$ differs slightly (see, Fig. 2). 

Using the solutions of equations with potential $V_+$ and $V_-$, 
the radial wave functions $R_{+ {1 \over 2}}$ and $R_{- {1 \over 2}}$ for
spin up and spin down particles respectively of the original Dirac equation are given below,
$$
{\rm Re}\left(R_{\frac{1}{2}} \Delta^{\frac{1}{2}}\right) = 
\frac{a_+ {\rm cos}(u - \theta) +
a_- {\rm cos}(u + \theta)}{2\sqrt{k}} + \frac{a'_+ {\rm cos}(u' - \theta) 
- a'_- {\rm cos}(u' + \theta)}
{2\sqrt{k'}}
\eqno{(23a)}
$$
$$
{\rm Im}\left(R_{\frac{1}{2}} \Delta^{\frac{1}{2}}\right) = 
\frac{a_+ {\rm sin}(u - \theta) -
a_- {\rm sin}(u + \theta)}{2\sqrt{k}} + \frac{a'_+ {\rm sin}(u' - \theta) 
+ a'_- {\rm sin}(u' + \theta)}{2\sqrt{k'}}
\eqno{(23b)}
$$
$$
{\rm Re}\left(R_{-\frac{1}{2}}\right) = \frac{a_+ {\rm cos}(u + \theta) +
a_- {\rm cos}(u - \theta)}{2\sqrt{k}} - \frac{a'_+ {\rm cos}(u' + \theta) 
- a'_- {\rm cos}(u' - \theta)}
{2\sqrt{k'}}
\eqno{(23c)}
$$
$$
{\rm Im}\left(R_{-\frac{1}{2}}\right) = \frac{a_+ {\rm sin}(u + \theta) -
a_- {\rm sin}(u - \theta)}{2\sqrt{k}} - \frac{a'_+ {\rm sin}(u' + \theta) 
+ a'_- {\rm sin}(u' - \theta)}
{2\sqrt{k'}}
\eqno{(23d)}
$$
Here, $a_+=(A_+-A_{+h})/\sqrt(q/k)$ and $a_-=(A_--A_{-h})/\sqrt(q/k)$.
Here, we have brought back $k$ and $k'$  so that these may resemble the
original solution (eq. 16)  using WKB approximation.
$\frac{a'_+}{\sqrt{k'}}$ and $\frac{a'_-}{\sqrt{k'}}$ are the transmitted
and reflected amplitudes respectively for the wave of corresponding potential $V_-$.

Figure 5 shows the resulting wave functions for both the spin $+\frac{1}{2}$
and spin $-\frac{1}{2}$ particles respectively. The eiconals used
in plotting these functions (see, eq. 23[a-d]) have been calculated 
by approximating $V_\pm$ in terms of polynomials (This was done since $V_\pm$ 
as presented in eq. 10 is not directly integrable.) and using the 
definition $u ({\hat r}_*) = \int \sqrt(\sigma^2-V_\pm) d {\hat r}_* $. 
Note that the amplitude as well as the  wavelength remain constants in 
regions where $k$ is also a constant. As discussed before,
the wave functions are almost sinusoidal close to the horizon and
at a very large distance (albeit with different frequencies). Since the
net current ($|P_{+\frac{1}{2}}|^2-|P_{-\frac{1}{2}}|^2$) is conserved, 
probability of spin $+\frac{1}{2}$ is complimentary to the probability 
of spin $-\frac{1}{2}$ particles respectively. 

\subsection*{Solutions of Region II}
Here we study the second region where for any set of physical
parameter total energy of the incoming particle is less than the maximum
height of the potential barrier. Thus, the WKB approximation is not valid 
in the whole range of $\hat{r}_*$. In such regions, the solutions will
be a linear combination of Airy functions because the
potential is approximately linear in ${\hat{r}}_*$ in those
intervals. At the junctions one has to match the solutions
with Airy functions along with the solution obtained by the WKB method.
In the region where the WKB approximation is valid, local values 
of reflection and transmission coefficients and the wave functions
can be calculated easily by following the same method described 
in Case I. In other regions, the equation reduces to
$$
{{d^{2}Z_{+}} \over {d{\hat{r}}_{*}^{2}}} - x Z_{+} = 0,
\eqno{(24)}
$$
where, $x = \beta^{1 \over 3} ({\hat{r}}_{*} - p)$, $\beta$ is
chosen to be positive and $p$ is the critical
point where the total energy and potential energy are matching.

Let $Z_{+}(x) = x^{1 \over 2} Y(x)$
and considering region $x > 0$ the equation (24) reduces to
$$
x^{2} {{d^{2}Y} \over {dx^{2}}} + x {{dY} \over {dx}} - \left(x^{3} +
{1 \over 4}\right) Y(x) = 0 .
\eqno{(25)}
$$
By making yet another transformation,
$$
\xi = {2 \over 3} x^{3 \over 2} ,
\eqno{(26)}
$$
we obtain,
$$
\xi^{2} {{d^{2}Y} \over {d\xi^{2}}} + \xi {{dY} \over {d\xi}} -
\left(\xi^{2} + {1\over 9}\right) Y(\xi) = 0 .
\eqno{(27)}
$$
This is the modified Bessel equation.
The solution of this equation is $I_{+ {1 \over 3}}(\xi)$ and $I_{- {1 \over 3}}(\xi)$.
Hence, the solution of eq. (27) will be,
$$
Z_{+}(x) = x^{1 \over 2} [C_{1} I_{+ {1 \over 3}}(\xi) + C_{2} I_{- {1 \over 3}}(\xi)].
\eqno{(28)}
$$

When $x<0$ the corresponding equation is,
$$
\xi^{2} {{d^{2}Y} \over {d\xi^{2}}} + \xi {{dY} \over {d\xi}}
+ \left(\xi^{2} - {1 \over 9}\right) Y(\xi) = 0,
\eqno{(29)}
$$
which is the Bessel equation. The corresponding solution is
$$
Z_{+}(x) = | x |^{1 \over 2} [D_{1} J_{+ {1 \over 3}}(\xi) + 
D_{2} J_{- {1 \over 3}}(\xi)],
\eqno{(30)}
$$
where $J_{\pm}$ and $I_{\pm}$ are the Bessel functions and the modified Bessel
functions of order $\frac{1}{3}$ respectively.

The Airy functions are defined as
$$
Ai(x) = {1 \over 3} x^{1 \over 2} [I_{- {1 \over 3}}(\xi) - I_{+ {1 \over
3}}(\xi)],  \hskip1cm x > 0 ,
\eqno{(31)}
$$

$$
Ai(x) = {1 \over 3} | x |^{1 \over 2} [J_{- {1 \over 3}}(\xi) + J_{+ {1 \over
3}}(\xi)], \hskip1cm x < 0 ,
\eqno{(32)}
$$

$$Bi(x) = {1 \over \surd{3}} x^{1 \over 2} [I_{- {1 \over 3}}(\xi) + I_{+ {1
\over 3}}(\xi)], \hskip1cm x > 0 ,
\eqno{(33)}
$$

$$
Bi(x) = {1 \over \surd{3}} | x |^{1 \over 2} [J_{- {1 \over 3}}(\xi) - J_{+ {1
\over 3}}(\xi)], \hskip1cm  x < 0 .
\eqno{(34)}
$$

In terms of Airy functions, the solutions (28) and (30) can be written as
$$
Z_{+} = {3 \over 2} (C_{2} - C_{1}) Ai(x) + {\surd{3} \over 2} (C_{2} + C_{1}) Bi(x)
\hskip0.5cm {\rm for}\hskip0.2cm x > 0,
\eqno{(35)}
$$
$$Z_{+} = {3 \over 2} (D_{2} + D_{1}) Ai(x) + {\surd{3} \over 2} (D_{2} - D_{1}) Bi(x)
\hskip0.5cm {\rm for}\hskip0.2cm x < 0.
\eqno{(36)}
$$
By matching boundary conditions it is easy to show that the solution corresponding
$x > 0$ and that corresponding $x < 0$ are continuous
when $C_1 = - D_1$ and $C_2 = D_2$.

To have an explicit solution, we choose the following set of parameters:
$ M = 1$, $ m_p = 0.1$, $l =  {1 \over 2}$, $m = - {1 \over 2}$ and
$ \sigma = 0.15$.

In Fig. 6, we show the nature of $V_+$ and $V_-$. However, while solving, we use
the equation containing $V_{+}$ (eq. 9). Unlike the case in the
previous Section, here $\sigma^2$ is no longer
greater than $V_\pm$ at all radii. As  a result,
$k^2$ may attain negative values in some region.
In Fig. 7, nature of $V_{+}$ (solid curve), parameter $k$ (dashed curve) and
energy $E$ (short-dashed curve) are shown.
Here, WKB approximation can be applied  in regions other than
${{\hat{r}}_{*}}  \sim  - 6$ to $- 1$ and $4$ to $8$ where $k$ is close to zero
and the condition ${1 \over k} {dk \over d{{\hat{r}}_{*}}} < < k$ is not satisfied.
In  the region  ${\hat{r}}_{*}\sim 8$ to $4$ around the turning point 
${\hat{r}}_{*}\sim 5.6088$ the solutions turns out as [13]
$$
Z_{+} = 1.858386 Ai(x) + 0.600610914 Bi(x).
\eqno{(37)}
$$ 
Similarly, the solution from $- 1$ to $- 6$ i.e. around the turning point
${\hat{r}}_{*} = - 3.0675$ can be calculated as [13]
$$
Z_{+} = 1.978145 Ai(x) + 0.7168807 Bi(x).
\eqno{(38)}
$$
It is to be noted that in the region ${\hat{r}}_{*} \sim 4$ to $-1$, even though the
potential energy dominates over the total energy, WKB approximation method is
still valid. Here the solution will take the form
$\frac {exp(- u)} {\surd k}$ and $\frac {exp(+ u)} {\surd k}$.
Asymptotic values of the instantaneous reflection and the transmission coefficients
(which are traditionally known as the `reflection' and `transmission' coefficients
respectively) are obtained from the WKB approximation. This yields 
the integral constant $c$ as in Case I.
>From eq. 22(a-b) local reflection and transmission coefficients
are calculated, behaviour of which are shown in Fig. 8. The constants $A_{-h}$
and $A_{+h}$ are calculated as before. Note the decaying nature
of the reflection coefficient inside the potential barrier.

\section*{V. SOLUTION OF THE EQUATIONS BY STEP-POTENTIAL METHOD}

In the above sections we presented our semi-analytical solutions by WKB
method with an appropriate boundary condition at the horizon. 
A numerical approach would be to replace the
potential $V({\hat r}_*)$ by a collection of step function as shown in Fig. 9a.
Here, the solid steps approximate the dashed potential for $m_p=0.8$ and $\sigma=0.8$.
The standard junction conditions of the type,
$$
Z_{+, n}=Z_{+, n+1}
\eqno{(39a)}
$$
where,
$$
Z_{+,n}  = A_{n} exp [ik_n {\hat r}_{*, n}] + B_{n} exp [-ik_n{\hat r}_{*, n}]
$$
and
$$ 
\frac{dZ_{+}}{d{\hat r}_*}|_n = \frac{dZ_{+}}{d{\hat r}_*}|_{n+1}
\eqno{(39b)}
$$
where,
$$
\frac{dZ_{+}}{d{\hat r}_*}|_n  = ik_nA_n exp (ik_n {\hat r}_{*,n}) - ik_nB_n 
exp (-ik_n{\hat r}_{*,n})
$$
at each of the $n$ steps were used to connect solutions at successive steps. 
As before, we use the inner boundary condition, to be $R \rightarrow 0$ 
at ${\hat r}_* \rightarrow -\infty $. In reality,
we used as many as 12000 steps to accurately follow the shape of the
potential. Smaller step sizes were
used whenever $k$ varies faster. Fig. 9b shows the comparison of the 
instantaneous reflection coefficients in both the methods. The solid 
curve is from the WKB method of previous section and the dotted curve 
is from the step-potential method as we described here. The agreement is
clearly excellent.

\section*{VI. BLACK HOLE: A MASS SPECTROGRAPH?}

In order to show that the black hole scatters incoming waves of different
rest masses ($m_p$) and of different energies ($\sigma^2$)
quite differently, we show a collection of solutions in Figs. 10(a-d).
In Figs. 10a, we show reflection and transmission coefficients for
waves with parameters $\sigma = 0.8$ (solid), $0.85$ (dotted) and $0.90$
(dashed) respectively with the same $m_p=0.8$. As the energy of the particle rises
comparable to the height of the potential (which is solely dependent on
$m_p$ at a large distance), the reflection coefficient goes down and the
transmission coefficient goes up. In Fig. 10b, the real part of the wave $Z_+$,
corresponding  to these three cases are shown.
At ${\hat r}_*=0$, the wave pattern is independent of $\sigma$ as the phase factor
is trivially the same in all the cases. The dispersal of the wave
with frequency is clear. Waves with smaller energy and
longer wavelength are scattered with higher amplitude of 
Re($Z_+$) as the fraction of the
reflected wave goes up when energy is reduced. This behaviour is
valid till $R<0.5$ since the amplitude of Re($Z_+$)= $(1+\sqrt{TR})^{1/2}$. 
For $R>0.5$, amplitude of Re($Z_+$) goes down with energy.
In Fig. 10(c-d), solutions are shown with varying the rest mass of the particles while
keeping  $\sigma$ fixed at $0.8$. The solid curve, dotted curve and
the dashed curves are for $m_p=0.8$, $0.76$, $0.72$ respectively. Most interesting aspect
is that close to the black hole ${\hat r}_* \lsim 0$, the reflection and transmission
coefficients as well as the nature of the wave are quite independent of the
rest mass. This is understandable, as just outside the horizon, the potential is
insensitive to $m_p$. However, farther out, amplitude of 
Re($Z_+$) goes up as before when $m_p$ is raised as larger fraction of 
the wave is reflected from the potential (Fig.10c).

This interesting behaviour, for the first time shows that one could scatter
a mixed wave off a black hole and each of the constituent wave would behave
differently as in a prism or a mass spectrograph.

\section*{VII. CONCLUDING REMARKS}

In this paper, we analytically studied scattering of spin-half particles from
a Schwarzschild black hole. In particular, we presented the nature of the radial wave
functions and the reflection and transmission coefficients. Our main motivation 
was to give an analytical expression of the solution which could be 
useful for further study of interactions among Hawking
radiations, for instance. We verified that these analytical solutions
were indeed correct by explicitly solving the same set of equations numerically
using step-potential approach as described in Section V. We classified 
the entire parameter space in terms of the physical and unphysical regions
and the physical region was further classified into two regions,
depending on whether the particle `hits' the potential barrier or not.
We chose one illustrative example in each of the regions. We emphasize that
the most `interesting' region to study would be close to $m_p \sim \sigma$.
However we pointed out (Fig. 1b) that for $m_p \leq 0.3$, WKB solutions
cannot be trusted, and other methods (such as those using Airy functions)
must be employed.

We used the well known WKB approximation method as well as the step-potential method
of quantum mechanics
to obtain the spatial dependence of the coefficients of the wave function.
This in turn, allowed us to determine the reflection and transmission 
coefficients and the nature of wave functions.
The usual WKB method with constant coefficients and (almost) constant wave number
$k$ is successfully applied even when the coefficients and  wave number are not constant
everywhere. Solution from this `instantaneous' WKB (IWKB for short) method
agrees fully with that obtained from a purely classical numerical method
where the potential is replaced by a collection of steps (simple 
quantum mechanical approach).
The resulting wave forms satisfy the inner and the outer boundary
conditions. Our method of obtaining solutions should be
valid for any black hole geometry which are asymptotically flat so that
radial waves could be used at a large distance.
This way we ensure that the analytical solution is closer to the exact
solution. In Region II, in some regions, WKB method cannot be applied and
hence Airy function approach or our step-potential approach could be used.

In the literature, reflection and transmission coefficients are defined
at a single point. These definitions are meaningful only if the
potential sharply changes in a small region while studies are made from a large distance
from it. In the present case, however, the potential changes over a large distance
and we are studying in these regions as well.
Although we used the words `reflection' and `transmission' coefficients,
in this paper very loosely, our definitions are very rigorous and well
defined. These quantities are simply the instantaneous values. In our belief
these are more physical. The problem at hand is very
similar to the problem of reflection and transmission of acoustic
waves from a strucked string of non-constant density where reflection and transmission
occurs at each point.

Unlike in the case of a Kerr black hole, the solution of the
angular equation around a Schwarzschild black hole is independent of 
the azimuthal or meridional angles [5-7]. This is expected because of
symmetry of the space-time. However, otherwise, the nature of the reflection and
transmission coefficients were found to strongly distinguish solutions
of different rest masses and different energies as illustrated in Figs. 10(a-d).
For instance,  when the energy of the wave is increased for a given mass
of the particle, reflected component goes down. In regions where $R>0.5$,
Re($Z_+$) goes down with energy, but where $R<0.5$, Re($Z_+$) goes up with energy.
In any case, the wave scattered off are distinctly different.
In a way, therefore, black holes can act as a mass spectrograph! 
For instance a mixture of waves should be splitted into its components by the black hole. 
Our method is quite general and should be used to study outgoing Hawking radiations also.
This is beyond the scope of this paper and would be dealt with in future.

\vskip 1cm



\section*{FIGURE CAPTIONS}

\begin{figure}
\noindent Fig. 1a: Classification of the parameter space in terms of the
energy and rest mass of the particles. The physical region $\sigma>m_p$
is further classified in terms of whether the particle actually `hits'
the barrier or not.
\end{figure}

\begin{figure}
\noindent Fig. 1b: Contours of constant $w_{max}$=
max$(\frac{1}{k^2}\frac{dk}{d{\hat{r}}_*})$ 
are shown to indicate that generally $w<<1$ and therefore WKB approximation
is valid in most of the physical region. Labels indicate values of $w$.
\end{figure}

\begin{figure}
\noindent Fig. 2: Behaviour of potentials 
$V_+$ (solid curve) and $V_-$ (dashed curve), as a function of $\hat{r}_*$.
The parameters are chosen from Region I of Fig. 1.
\end{figure}

\begin{figure}
\noindent Fig. 3: Behaviour of $V_+$ (solid curve), $k$ 
(dashed curve), total energy $E$ (short-dashed curve),
as functions of $\hat{r}_*$.
\end{figure}

\begin{figure}
\noindent Fig. 4: Behaviour of local transmission ($T$, solid curve) and 
reflection ($R$, dashed curve) coefficients, as functions of $\hat{r}_*$. 
Close to the horizon, transmission is a hundred percent
and reflection is almost zero.
\end{figure}

\begin{figure}
\noindent Fig. 5: Behaviour of (a) Re$\left(R_{1/2}\Delta^{1/2}\right)$,
(b) Im$\left(R_{1/2}\Delta^{1/2}\right)$, (c) Re$\left(R_{-1/2}\right)$, 
(d) Im$\left(R_{-1/2}\right)$, as a function of $\hat{r}_*$. Note the
complimentary nature wave functions of the spin $+\frac{1}{2}$ and spin 
$-\frac{1}{2}$ particles. This is because the current is conserved.
\end{figure}

\begin{figure}
\noindent Fig. 6: Behaviour of $V_+$ (solid curve) and $V_-$ (dashed curve),
as a function of $\hat{r}_*$. The parameters are chosen 
from Region II of Fig. 1(a-b).
\end{figure}

\begin{figure}
\noindent Fig. 7: Descriptions are same as in Fig. 3. See text for
the choice of parameters. 
\end{figure}

\begin{figure}
\noindent Fig. 8: Descriptions are same as in Fig. 4. See text for 
the choice of parameters.
\end{figure}

\begin{figure}
\noindent Fig. 9a: Steps (solid) approximating a potential (dotted) thus reducing the
problem to that of a quantum mechanics. The parameters are
$m_p=0.8$ and $\sigma=0.8$.
\end{figure}
\noindent Fig. 9b: Comparison of variation of instantaneous reflection
coefficient $R$ with the radial coordinate ${\hat r}_*$ using analytical
WKB method (solid) and numerical step-potential method (dotted). The
parameters are $m_p=0.8$ and $\sigma=0.8$.

\begin{figure}
\noindent Fig. 10(a-d): Comparison of (a) reflection and transmission coefficients
and (b) real amplitude of the wave function $Z_+$ for $m_p=0.8$ and $\sigma=0.80$
(solid), $0.85$ (dotted) and $0.90$ (dashed) respectively. (c-d) Similar
quantities for $m_p=0.80$, (solid) $0.76$ (dotted) and $0.72$ (dashed)
respectively keeping $\sigma=0.8$ fixed. Higher reflection component enhances the
wave amplitude, thus differentiating the incoming waves very clearly.
\end{figure}

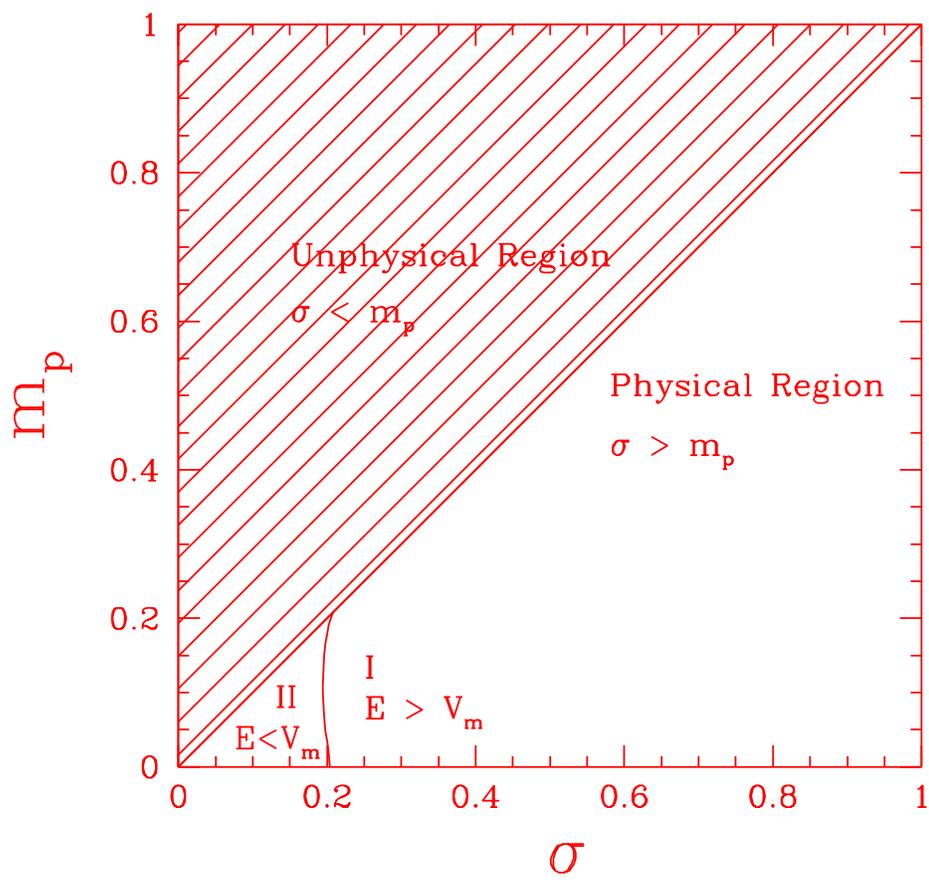
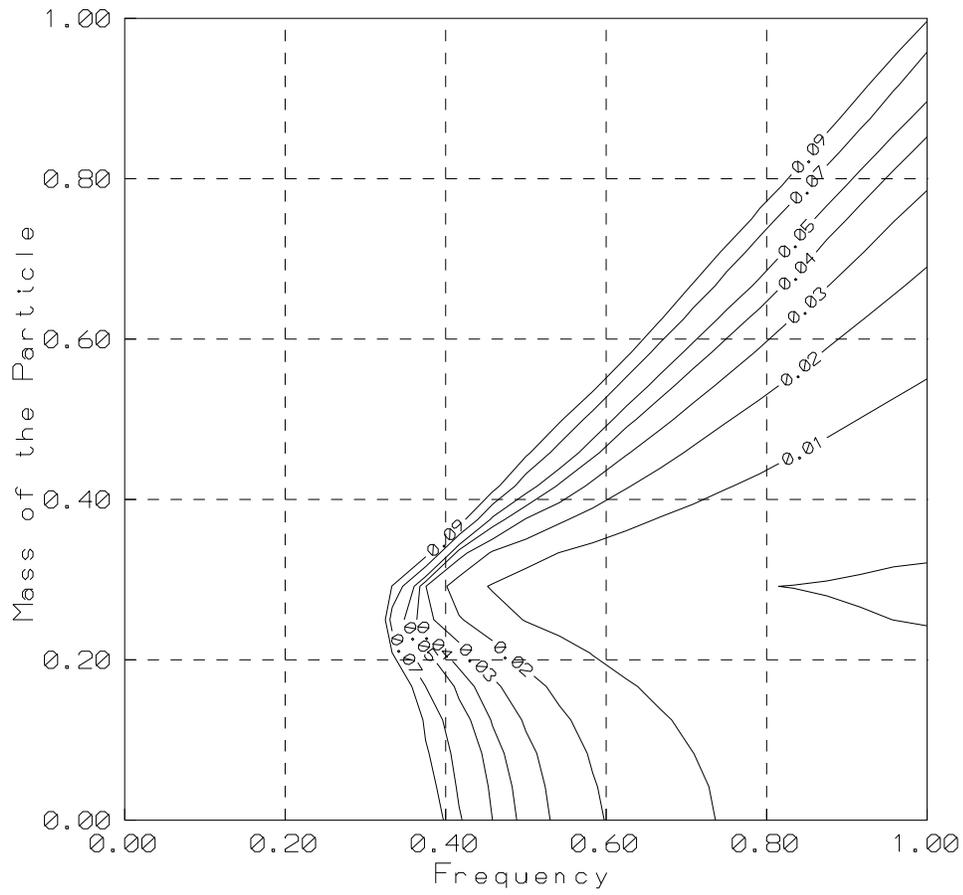
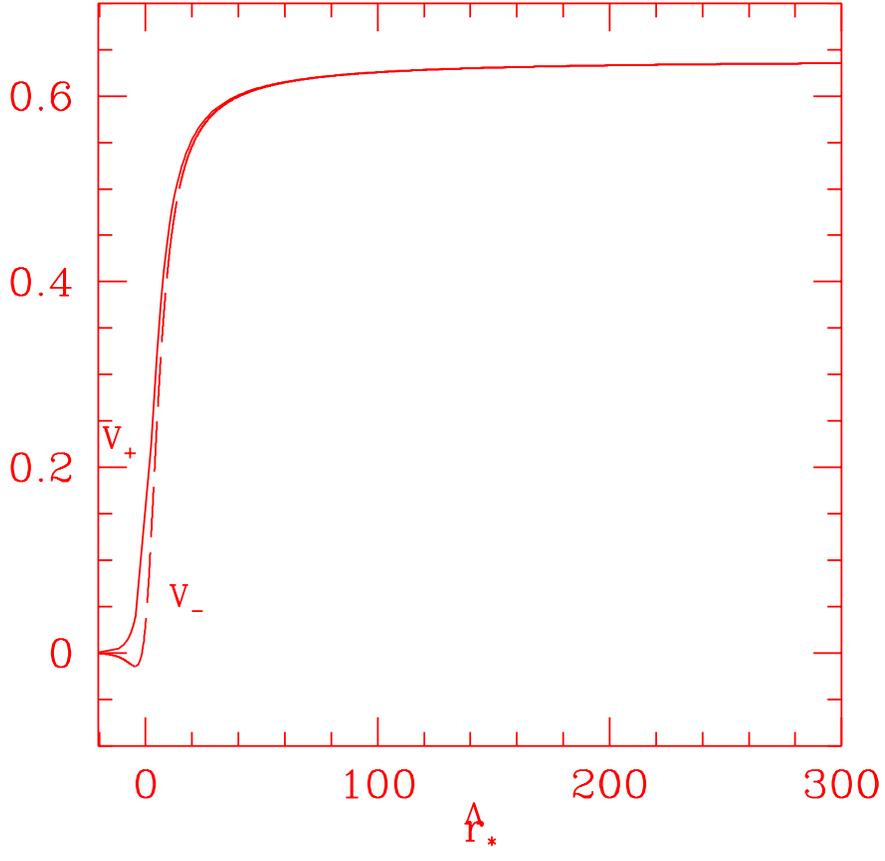
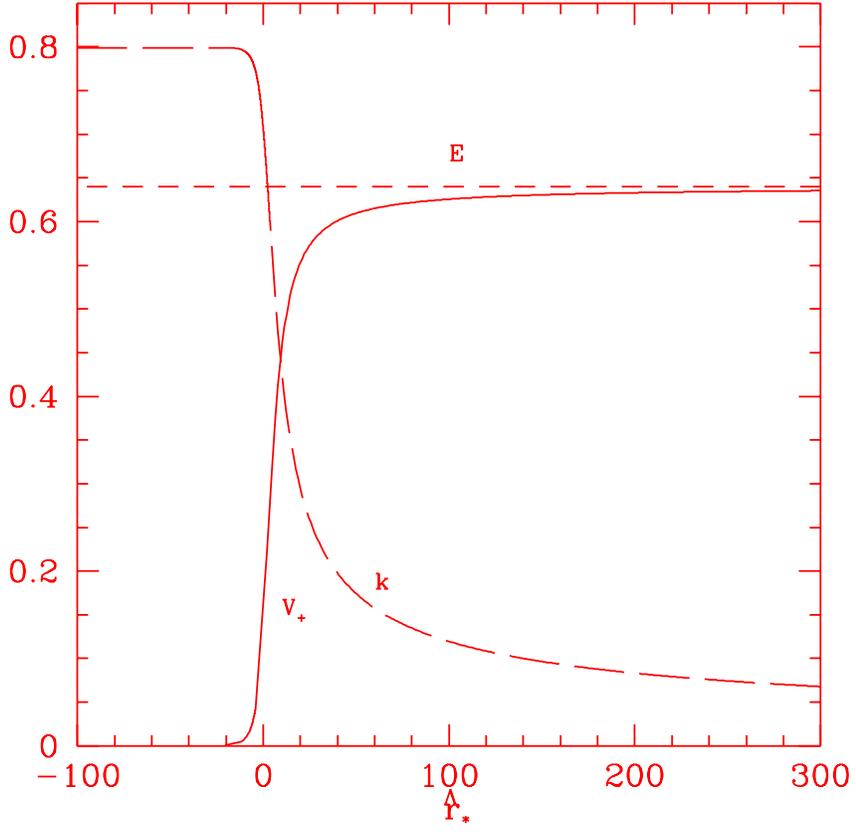
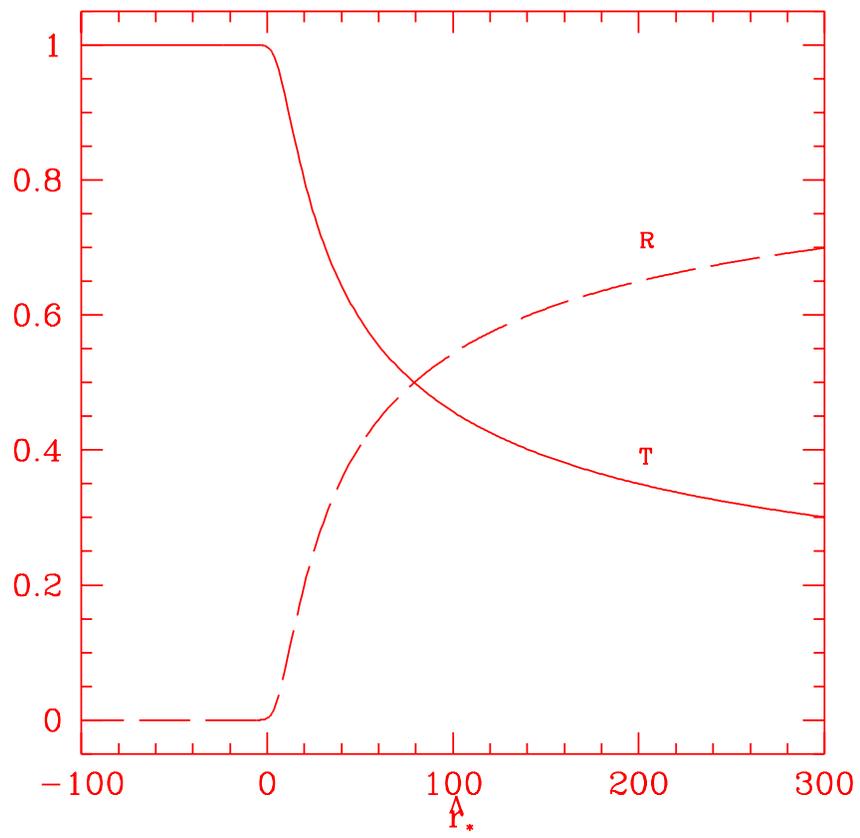
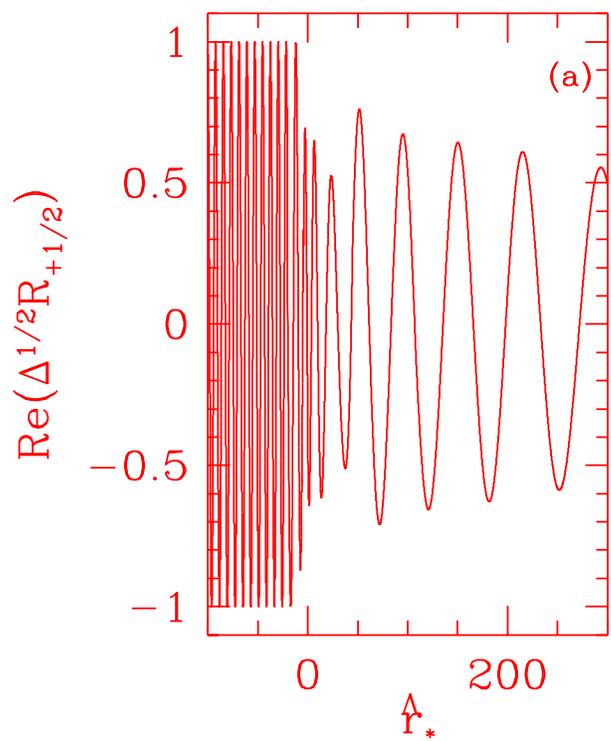
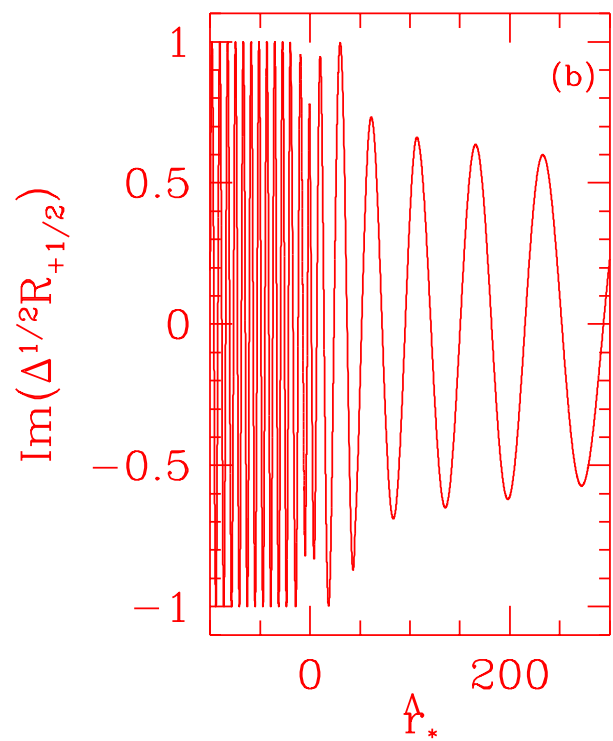
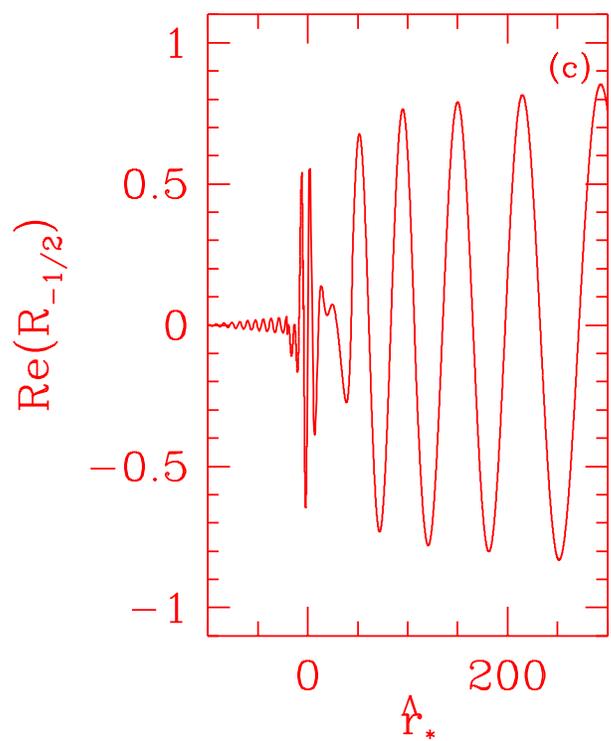
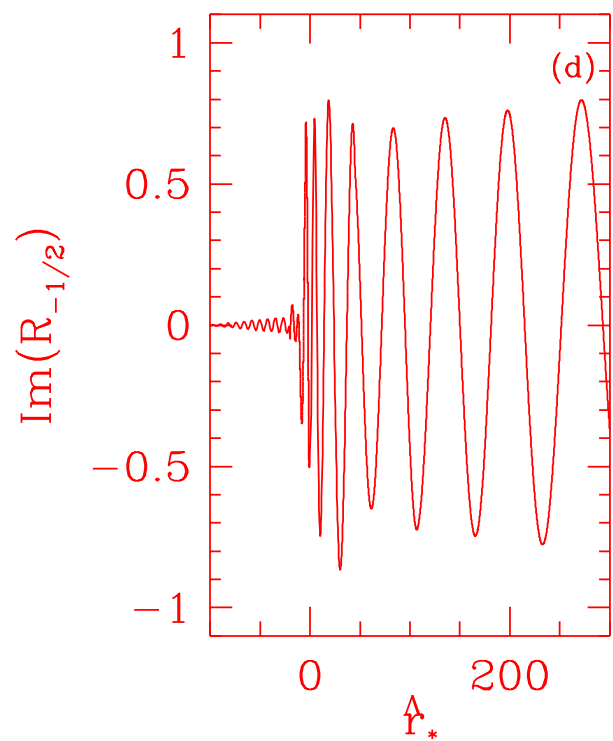
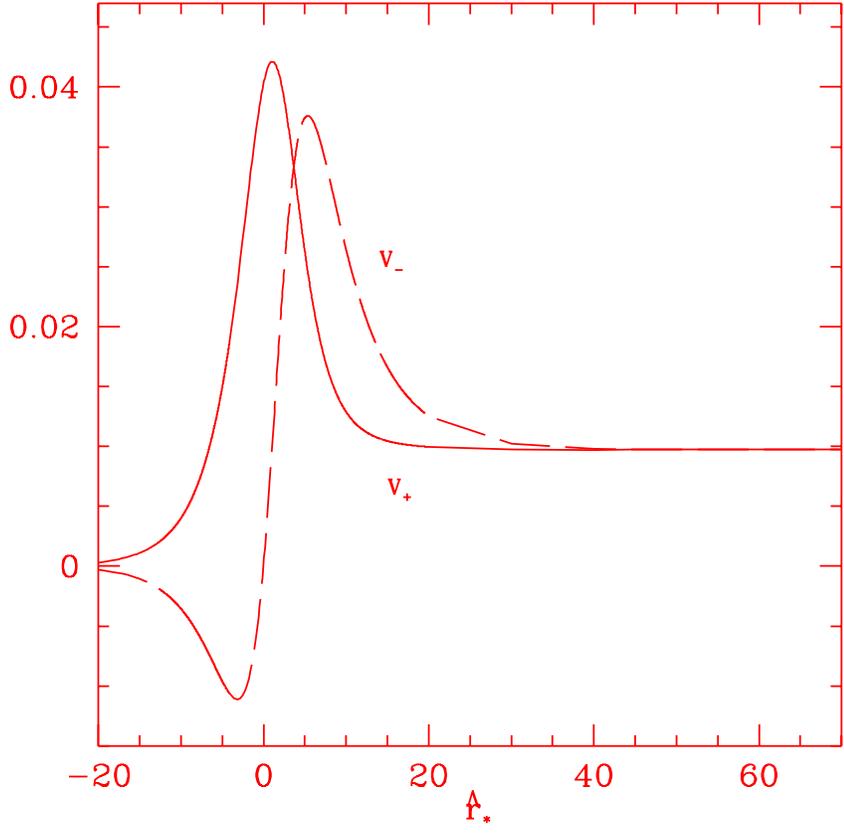
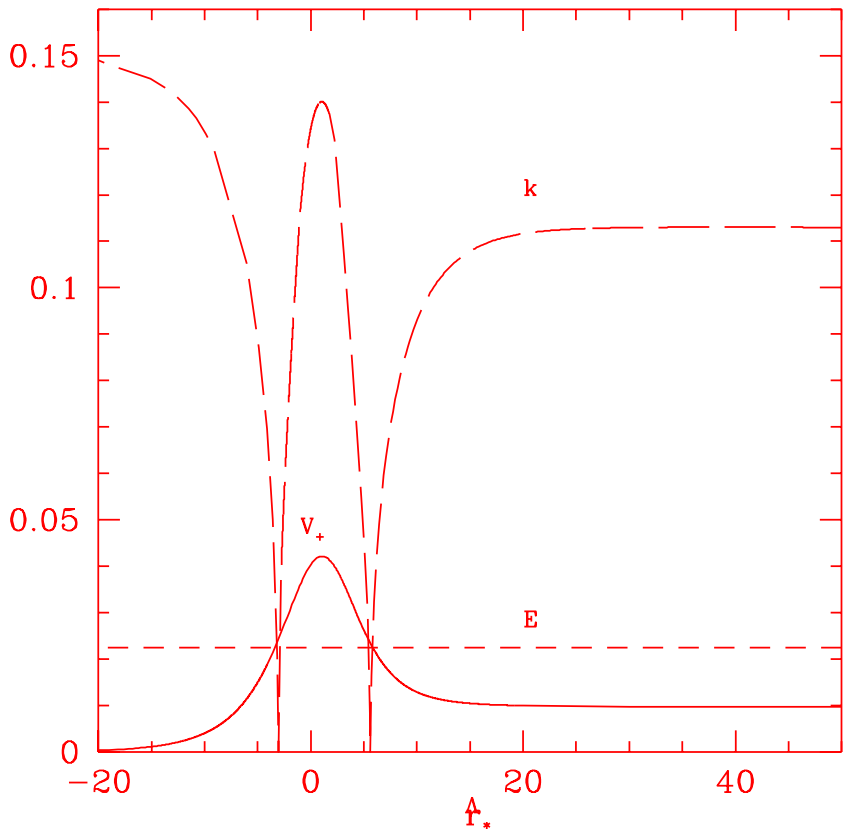
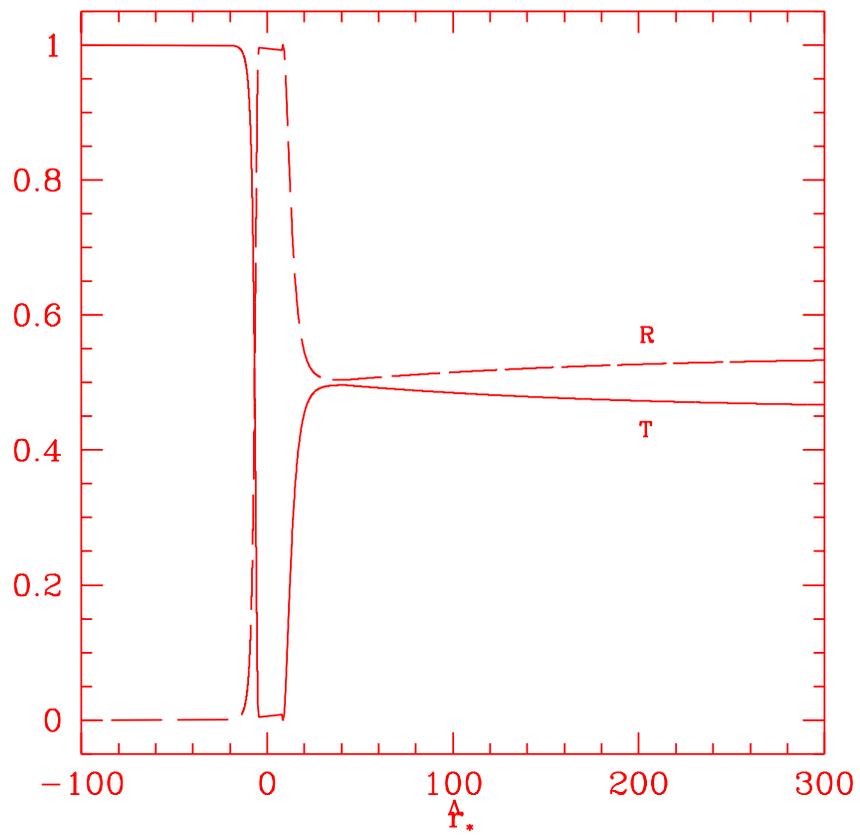
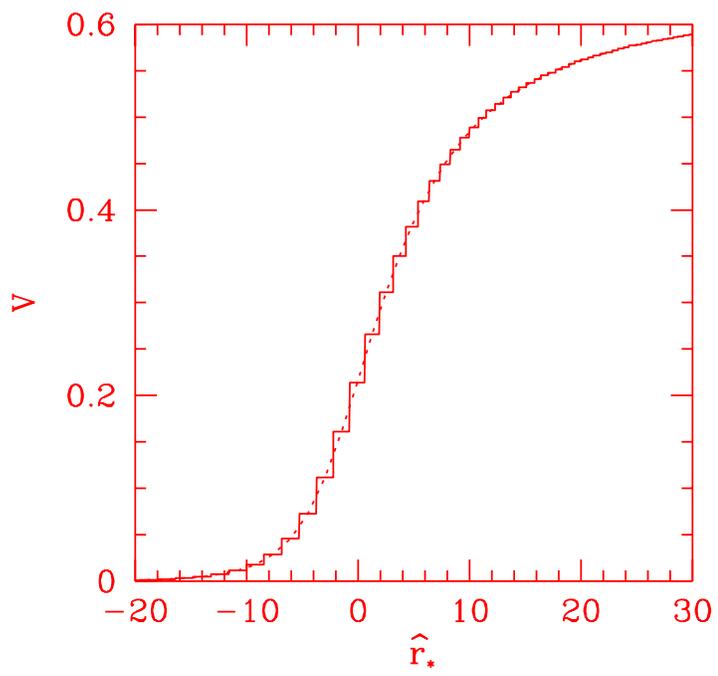
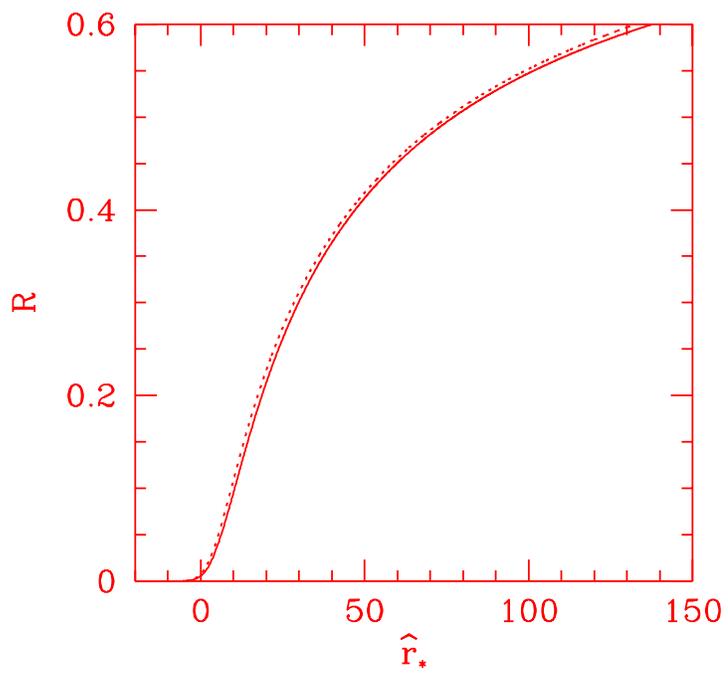
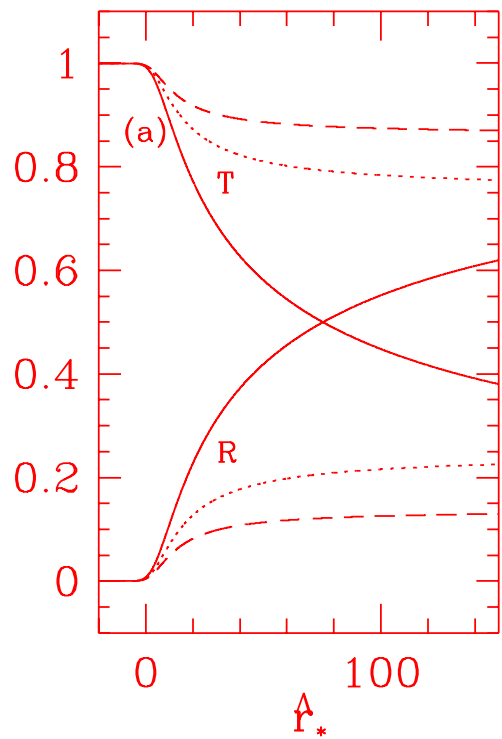
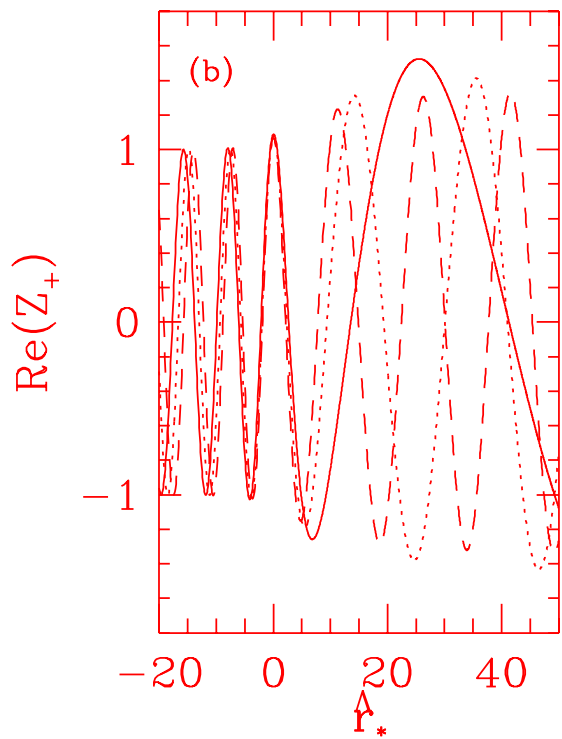
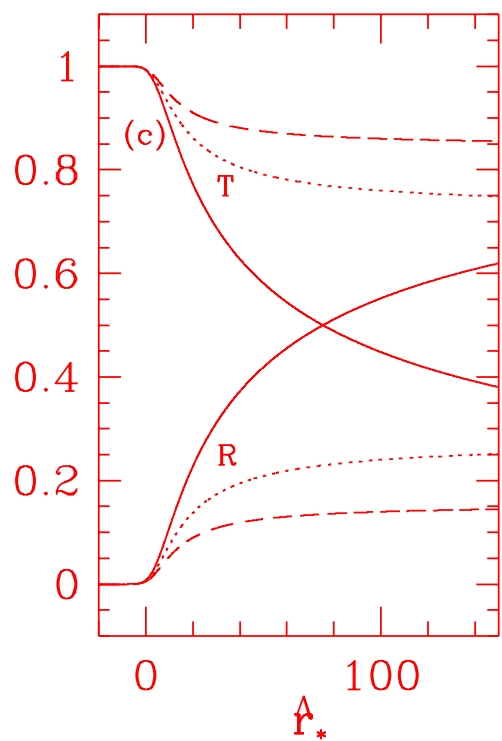
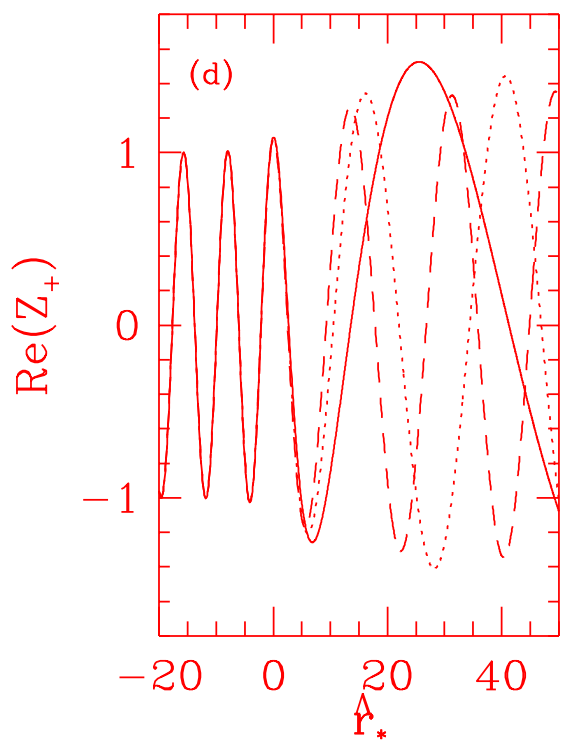
\end{document}